\documentclass[twocolumn,twocolumn]{IEEEtran}
\usepackage[T1]{fontenc}
\usepackage{color}
\usepackage{float}
\usepackage{mathrsfs}
\usepackage{amsmath}
\usepackage{amssymb}
\usepackage{graphicx}
\usepackage[unicode=true,
 bookmarks=true,bookmarksnumbered=true,bookmarksopen=true,bookmarksopenlevel=1,
 breaklinks=false,pdfborder={0 0 0},pdfborderstyle={},backref=false,colorlinks=false]
 {hyperref}
\hypersetup{pdftitle={Your Title},
 pdfauthor={Guo yafei},
 pdfpagelayout=OneColumn, pdfnewwindow=true, pdfstartview=XYZ, plainpages=false}

\makeatletter

\floatstyle{ruled}
\newfloat{algorithm}{tbp}{loa}
\providecommand{\algorithmname}{Algorithm}
\floatname{algorithm}{\protect\algorithmname}

\usepackage[caption=false,font=footnotesize]{subfig}
\usepackage{algorithm}
\usepackage{algorithmic}

\usepackage{multirow} %multirow for format of table 
\usepackage{amsmath} 
\usepackage{xcolor}

\allowdisplaybreaks[4]

\ifCLASSOPTIONcompsoc
\usepackage[caption=false,font=normalsize,labelfont=sf,textfont=sf]{subfig}
\else
\usepackage[caption=false,font=footnotesize]{subfig}
\fi

\usepackage{cite}
\usepackage{bm}
\usepackage{algorithmic}
\usepackage{algorithm}
\usepackage{graphicx}
\IEEEoverridecommandlockouts
\usepackage{amsthm}
\usepackage{lettrine}

\usepackage{hyperref}

\usepackage{geometry}
\@ifundefined{showcaptionsetup}{}{%
 \PassOptionsToPackage{caption=false}{subfig}}
\usepackage{subfig}
\makeatother

\begin{document}

\title{\textcolor{black}{UAV-Assisted MEC for Disaster Response: Stackelberg Game-Based Resource Optimization}}

\author{\IEEEauthorblockN{Yafei Guo$^{1}$, Ziye Jia$^{1}$, Lei Zhang$^{1}$, Jia He$^{1}$, Yu Zhang$^{2}$, and Qihui Wu$^{1}$\\
}
\IEEEauthorblockA{
\small
$^{1}$The Key Laboratory of Dynamic Cognitive System of Electromagnetic Spectrum Space, Ministry of Industry and\\
Information Technology, Nanjing University of Aeronautics and Astronautics, Nanjing, Jiangsu, 211106, China\\
$^{2}$Department of Electronic and Electrical Engineering, University College London, London WC1E 6BT, United Kingdom\\
\{guoyafei, jiaziye, Zhang\_lei, 071940128hejia, wuqihui\}@nuaa.edu.cn, uceezhd@ucl.ac.uk}
% \IEEEauthorblockA{
% $^{\dagger}$Key Laboratory of Dynamic Cognitive System of Electromagnetic Spectrum Space, Ministry of Industry and\\ 
% Information Technology, Nanjing University of Aeronautics and  Astronautics, Nanjing, Jiangsu, 211106, China\\
% $^{\S}$Department of Electronic and Electrical Engineering, University College London, London WC1E 6BT, United Kingdom\\
% \{hanyulu, jiaziye, sx2304109, wuqihui\}@nuaa.edu.cn, uceezhd@ucl.ac.uk
% }
\thanks{{This work was supported in part by National Natural Science Foundation of China under Grant 62301251, 
in part by the Natural Science Foundation on Frontier Leading Technology Basic Research Project of Jiangsu under Grant BK20222001,  
in part by the Aeronautical Science Foundation of China 2023Z071052007,
and  in part by the Young Elite Scientists Sponsorship Program by CAST 2023QNRC001.
 }}
}
\maketitle

\thispagestyle{empty}
\begin{abstract}
    The unmanned aerial vehicle assisted multi-access edge computing (UAV-MEC) technology has been widely applied in the sixth-generation era. 
    However, due to the limitations of energy and computing resources in disaster areas, how to efficiently offload the tasks of damaged user equipments (UEs) to UAVs is a key issue.
    In this work, we consider a multiple UAV-MECs assisted task offloading scenario,  which is deployed inside the three-dimensional corridors and provide computation services for UEs.
    In detail, a ground UAV controller acts as the central decision-making unit for deploying the UAV-MECs and allocates the computational resources.
    Then, we model the relationship between the UAV controller and UEs based on the Stackelberg game.
    The problem is formulated to maximize the utility of both the UAV controller and UEs.
    To tackle the problem, we design a K-means based UAV localization and availability response mechanism to pre-deploy the UAV-MECs.
    Then, a chess-like particle swarm optimization probability based strategy selection learning optimization algorithm is proposed to deal with the resource allocation.
    Finally, extensive simulation results verify that the proposed scheme can significantly improve the utility of the UAV controller and UEs in various scenarios compared with baseline schemes.
\end{abstract}
\begin{IEEEkeywords}
    UAV-assisted MEC, Stackelberg game, improved particle swarm optimization algorithm. 
    \end{IEEEkeywords}

\newcommand{\CLASSINPUTtoptextmargin}{0.8in}

\newcommand{\CLASSINPUTbottomtextmargin}{1in}

\section{Introduction}
\lettrine[lines=2]
W{ith} the rapid development of sixth generation communications, multi-access edge computing (MEC) technique is raised to provide computing services for various user equipments (UEs)\cite{you}.%\cite{you,1}.
However, for the UEs in the disaster-stricken area, the ground base stations (BSs) with MEC devices are damaged and unable to provide timely services.
Due to the high flexibility and mobility, UAV-MEC can serve as aerial BSs to provide timely and efficient services\cite{chess,xin2}.
However, there exist security issues and resource limitations for the aerial deployment of UAV-MEC.
% LoRa is a type of LPWAN and has been widely used in various iot applications due to its remote capabilities and low power requirements\cite{lora1}.
% LoRa With its special Chirp Spread Spectrum modulation(CSS), the longest reliable range of LoRa 2.4GHz is between 1700 and 2500 meters\cite{lora2}.
% UAVs are considered as a promising technique for area coverage due to their flexibility and adaptability.
% The hovering point selection of UAV-MEC is one of the key factors that affect the energy consumption in UAV-assisted MEC model\cite{jia2}.

There are several recent works related to UAVs that address the above issues.
% With the surge in the number of UAVs, low-altitude scenarios will be complicated, and low-altitude safety will be particularly important.
% ASTM has published their standards for Remote ID, including network type and wide type.
% Air corridors defined by appropriate air traffic regulators are also one of the means to ensure airspace safety\cite{way1}\cite{way2}.
In \cite{UAV-MEC1}, the optimizing of three-dimensional (3D) trajectory of UAV-MEC is considered to ensure the safety.
The 3D air corridor defines the flight path of UAVs, reducing collision risks and adding integrations with local airspace management systems\cite{way3}.
% In \cite{UAV-MEC2}, the power allocation of UAV-MEC is optimized to improve the throughput and energy efficiency of the system.
The remote identification (Remote ID) is a new communication system for ensure flight safety by obtaining the information of the UAV\cite{RID1}.
% In \cite{cui}, the power consumption was reduced by optimizing the access scheme and power allocation in hierarchical UAV-MEC.
In \cite{xin3}, a cooperative cognitive dynamic system is proposed to optimize the management of UAVs.
In addition, for the energy consumption of UAV-MEC, since UAV-MEC has a competitive relationship with UEs, the game theory is an effective method to solve such problems\cite{5}.
For instance, authors in \cite{6} utilize the Stackelberg game to maximize the profit of the UAV considering the delay, energy consumption and urgency.
% In \cite{7}, the game relationship between the ground BS and the UE is established.
The Potential game is proposed to solve the optimization problem for UAV trajectory planning, resource management and task offloading strategy in \cite{8}.
However, these works lack unified considerations of the safety and efficiency of UAV-MEC.

In this work, we consider a task offloading scenario, where the UAV-MECs are deployed in the 3D corridor and provide computation services for damaged UEs. 
% The UAV controller act as the central decision-making unit implements UAV-MEC deployments and computing resource allocations strategy using the Remote ID system. 
A ground UAV controller serves as the central decision-making unit, implementing UAV-MEC deployment and resource allocation strategies through the Remote ID.
% In order to maximize the utility of the UAV controller and UEs,
% the Stackelberg game is adopted to depict the interaction process between UEs and the UAV controller server. 
To maximize the utilities, the UEs and the UAV controller are formulated as a Stackelberg game problem.
% and the optimal relationship between resource price and amount of offloaded data was established.
Then, we design a UAV localization and availability response mechanism (ULAR) based on K-means to pre-deploy the UAVs and enable more efficient use of computing resources.
Further, we propose a chess-like particle swarm optimiza-tion probability-based strategy selection learning optimization algorithm (CPPO) to deal with the resource allocation.
Finally, we conduct extensive simulations to verify the effectiveness of the proposed methods.

This paper is organized as follows. 
In Section II, we present the system model and problem formulation. 
% multiple UAV-MEC assisted task offloading scenario 
% Section III address the task offloading incentive problem.
% Section IV proposes UAV pre-deployment algorithm and UEs task offloading optimization algorithm.
Section III analyzes the optimization problem.
Algorithms are designed in Section IV.
Section V conducts simulations and analyzes the results. 
Finally, conclusions are drawn in Section VI.
% \textcolor{black}{
% Additionally, we employ a data-driven approach, utilizing a recurrent 
% neural network (RNN) to adaptively adjust the process noise and observation noise 
% matrices during the Kalman filtering process. By learning the dynamic 
% characteristics of the system from the data, we enable the IMM-KF to better 
% adapt to the nonlinear changes in the system, thereby enhancing the accuracy 
% and robustness of state estimation.}

\section{System Model And Problem Formulation\label{sec:System-Model}}
\subsection{Network Model}
\begin{figure}[t]
\centering
\includegraphics[scale=0.25]{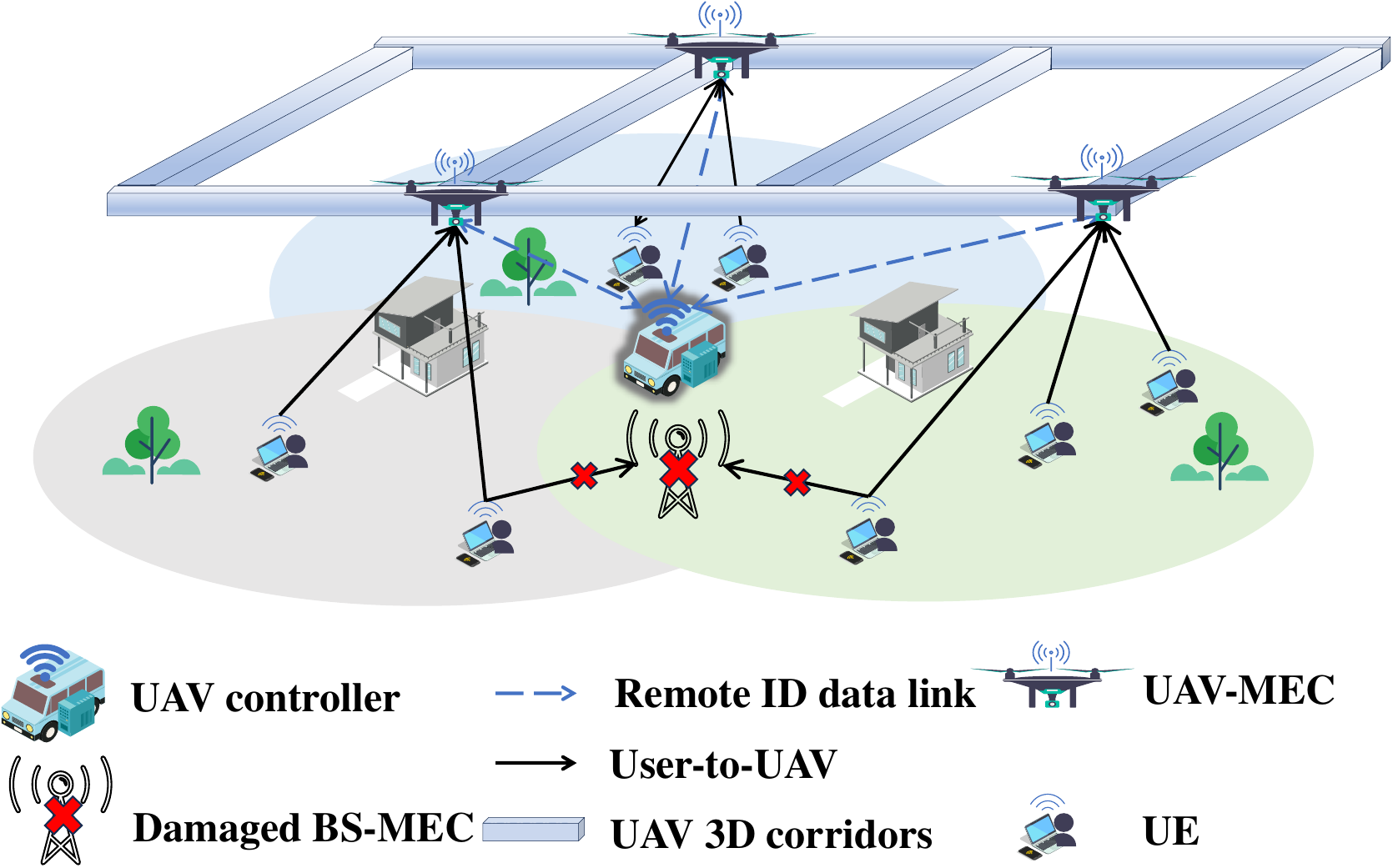}
\caption{\label{network} Multiple UAV-MECs assisted task offloading network. }
\end{figure}

As shown in Fig. \ref{network}, we consider a UAV-MEC assisted task offloading scenario, where UAVs are deployed at the 3D corridors\cite{way1,xin1}, 
and provide computing services for UEs in place of the Damaged BS-MEC that has lost communication and computing power.
In order to achieve the efficient collaborative operation of UAVs and MEC, the UAV controller deploys UAV-MEC through the precise Remote ID communication technology.
In detail, all $J$ UAVs denoted as $\mathcal{J}=\{1,2,\dots,J\}$ are dispatched by the UAV controller.
Further, $I$ UEs are distributed in the disaster area denoted as $\mathcal{I}=\{1,2,\dots,I\}$, 
and the computation task generated by UE $i$ is represented as a two-tuple, $T_i=\{G_i,g_i\}$. 
Wherein, $G_i$ and $g_i$ are the size of all data as well as the size of data offloaded to the UAV-MEC, respectively.
Let $\{p_i,\varepsilon_{i}\}$ denote the local computing power and unit energy consumption of UE $i$, respectively. 
The characteristics of UAV $j$ are described as $\{f_{j},P_{j}^\mathrm{comp}\}$, where $f_{j}$ represents the total computation resources owned by UAV $j$, 
and $P_{j}^\mathrm{comp}$ represents the CPU power of the UAV-MEC server.

The 3D Cartesian coordinates are utilized to represent the positions of UAVs and UEs.  
The location of UAV $j$ and UE $i$ are denoted as $z_{j}=(x_{j}, y_{j}, h_{j})$ and $z_{i}=(x_{i},y_{i},0)$, respectively.  
Thus, the distance between UAV $j$ and UE $i$ is claculated as $d_{i,j}=\sqrt{\left\|z_i-z_j\right\|^2}$.
% Remote ID is used to identify and authenticate UAVs based on their distinctive ID.
% It is relatively new compared to ADS-B and ADS-B-like systems.
% Therefore, the position of the UAV can be swiftly determined in accordance with the Remote ID protocol.
% LoRaWAN serves as an optimal solution for utilizing the Remote ID protocol to facilitate information transmission across various communication technologies.
% LoRaWAN is a cost-effective Internet telecommunications solution; it is also acknowledged as an economical remote technology.
% It can cover a range of 15 to 30 kilometers.

\vspace{-2mm}
\subsection{Communication Model}\label{sec:model}
Considering that the UAVs are deployed in the 3D corridors with a constant height of $H$, 
we assume that the task offloading link between UEs and UAVs can be modeled as a line-of-sight (LoS). 
The uplink model from UE $i$ to UAV $j$ is expressed as:
\begin{equation}
    r_{i,j}=B\log_2\left(1+\frac{q_ih_{i,j}}{\sigma^2}\right),
\end{equation}
where $B$ is the bandwidth allocated by UE $i$ to UAV $j$, $\sigma^2$ represents the noise power and $q_i$ is the transmission power of UE $i$.
The channel gain can be expressed as $h_{i,j}=d_{i,j}^{-\rho}$,
where $\rho$ denotes the path loss factor between UEs and UAVs.
\subsection{Utility of UE and UAV Controller}
\subsubsection{Utility of UE}
The uplink transmission time from UE $i$ to UAV $j$ can be calculated as $t_{i,j}^{\mathrm{trans}}=\frac{g_{i}}{r_{i,j}}$.
Then, the uplink transmission energy is $E_{i,j}^{\mathrm{trans}}=p_it_{i,j}^{\mathrm{trans}}$.
The local computation energy consumption of UE $i$ is
$E_{\mathrm{local}}^{\mathrm{comp}}=\varepsilon_{i}(G_{i}-g_{i})$.
Therefore, the utility of UE $i$ is
\begin{equation}
    \begin{split}
    U_i&=S(g_i)\delta_i-E_{i,j}^{\mathrm{trans}}-E_{\mathrm{local}}^{\mathrm{comp}}-\lambda_ig_i,\label{UUi}
    % &=\mathrm{ln}(1+g_i)\delta_{i}-p_{i}\frac{g_{i}}{r_{i,j}}-\varepsilon_{i}(G_{i}-g_{i})-\lambda_{i}g_{i},
    \end{split}
\end{equation}
where $S(g_i)=\ln(1+g_{i})$ is the satisfaction function that reflects
the satisfaction degree of UE. $\delta_i$ is the control coefficient
to assess the impact of UE satisfaction on its utility.
$\lambda_i$ represents the price of each unit of data offloaded by the UAV-MEC server for each UE.

\subsubsection{Utility of UAV Controller}
% All UAVs are outfitted with Remote ID systems, enabling real-time monitoring of their locations.
% Each UAV is operated through a UAV controller terminal that is equipped with a Remote ID receiver.
In face of simultaneous offloading of multiple UEs, the UAV-MEC server adopts an equal allocation strategy of computing power\cite{M}.
We assume that there are $M_j$ UEs offloading tasks to the UAV $j$ simultaneously. 
The computing time of UAV $j$ for the offloading data of UE $i$ can be calculated as
% \begin{equation}
    $t_{i,j}^{\mathrm{comp}}=\frac{\alpha g_{i}}{f_{ij}}$,
% \end{equation}
where $f_{ij}={f_{j}/M_{j}}$ represents UE $i$ obtains the computing resource from UAV $j$, and $\alpha$ is a coefficient related to data encoding.

Therefore, the computing energy consumption of UAV-MEC server $j$ for UE $i$ is
    $E_{i,j}^{\mathrm{comp}}=P_j^{\mathrm{comp}}t_{i,j}^\mathrm{comp}$.
The energy consumption equation of UAV $j$ is 
    $E_{j}^{\mathrm{comp}}=\sum_{i=1}^{N}X_{i,j}E_{i,j}^{\mathrm{comp}}$, 
where variable $X_{i,j}\in\{0,1\}$ represents the link status between UE $i$ and UAV $j$.
When $X_{i,j}=1$, it indicates that UE $i$ has establishes a connection with UAV $j$, enabling the allocation of computing resources. 
Conversely, $X_{i,j}=0$ means there exists no connection between UE $i$ and UAV $j$.
The hovering energy consumption is denoted by
    $E_j^{\mathrm{hov}}={P_j^\text{hov}}/{\eta}$
, where $P_j^{\mathrm{hov}}$ indicates the minimum power for hovering, and $\eta$ represents the power efficiency\cite{hov1}.
% Thus, the total energy consumption of the UAV $j$ is calculated as
    % $E_{j}=E_{j}^{\mathrm{comp}}+ E_j^{\mathrm{hov}}$.
% The utility of the UAV controller is equal to its revenue minus expenditure. The revenue is the payment obtained from all UEs, and the expenditure is the energy cost of all UAVs. 
Therefore, the utility of the UAV controller can be calculated as
\begin{equation}
    \begin{split}
    U_{con}&=\sum_{i=1}^N\lambda_ig_i-\sum_{j=1}^{J}E_{j}^{\mathrm{comp}}-\sum_{j=1}^{J}E_j^{\mathrm{hov}}.\label{UUcon}\\
    % &=\sum_{i=1}^N\lambda_ilg_i-\sum_{i=1}^N\sum_{j=1}^JX_{i,j}\frac{\alpha g_i\cdot M_{\mathrm{uav},j}}{f_{\mathrm{uav},j}}P_j^{\mathrm{comp}}-\sum_{j=1}^J\frac{P_j^\text{hov}}{\eta}
    \end{split}
\end{equation}
% Specifically, $X_{i,j}=1$ indicates that the UAV
% controller establishes a communication link between them, so
% vehicle c can offload its subtasks to UAV $j$. While $x_{i,j}=0$
% means that there is no communication link between them, so
% UE $i$’s subtasks cannot be offloaded to UAV $j$.

\subsection{Problem Formulation}
The optimization goal is to maximize the utility of both the UAV controller and UEs. 
The optimization problem for UE $i$ is
\begin{alignat}{3}
    \mathbf{P0}\textrm{:}\;&\underset{g_i,\lambda_i,X_{i,j}}{\textrm{max}}  \quad U_{i},\notag\\
    &\mbox{s.t. }\sum_{j\in \mathcal{J}} X_{i,j}\leq1,\forall i\in \mathcal{N},\label{yueshu1}\\
    &g_i\in[0,G_i] ,\forall g_i\in\boldsymbol{g},\label{yueshu2}\\
    &\lambda_{i}\in[\lambda_i^{min},\lambda_i^{max}] ,\forall\lambda_{i}\in\boldsymbol\lambda,\label{yueshu3}\\
    &X_{i,j}\in\{0,1\},\forall i\in\mathcal{N},\forall j\in \mathcal{J},\label{yueshu4}
\end{alignat}
where constraint (\ref{yueshu1}) denotes that UE $i$ can only select one UAV from set $\mathcal{J}$ to offload tasks.
Constraint (\ref{yueshu2}) indicates that the data amount of tasks offloaded by UE $i$ is no more than the data amount of the entire task $T_i$.
Constraint (\ref{yueshu3}) indicates the resource price range of the UAV controller, 
where $\lambda_i^{min}$ and $\lambda_i^{max}$ are the minimum and maximum prices, respectively.
Constraint (\ref{yueshu4}) indicates that UE $i$ has the option to decide whether or not to offload the task to UAV $j$.

% Under the condition of the optimal data offloading strategy for UE $i$.
The optimization problem for the UAV controller is:
\begin{alignat}{2}
    \mathbf{P1}\textrm{:}\;&\underset{\boldsymbol{g},\boldsymbol{\lambda},X_{i,j}}{\textrm{max}}  \quad U_{con},\notag\\
    &\mbox{s.t. }E_{j}^{\mathrm{comp}}+ E_j^{\mathrm{hov}}\leq\varepsilon, \forall j\in \mathcal{J},\label{yueshu5}\\
    % &g_i\in[0,G_i] ,\forall g_i\in\boldsymbol{g},\label{yueshu6}\\
    % &\lambda_{i}^{min}\in[\lambda_i^{min},\lambda_i^{max}] ,\forall\lambda_{i}\in\boldsymbol\lambda,\label{yueshu7}
%    &\sum_{i=1}^Nlg_iX_{i,j}\leq f_j,\forall j\in\mathcal{J},\label{yueshu2}\\
    % &g_i\in[0,G_i], \forall i\in\mathcal{N}, 
    &(\ref{yueshu2}), (\ref{yueshu3}),\nonumber
\end{alignat}
% Constraint (\ref{yueshu2}) represents that the sum of computational resources allocated to all UEs by UAV $j$ should not exceed its total computation resources.
where constraint (\ref{yueshu5}) indicates that the energy consumption can not exceed the UAV battery budget.
% constraint (\ref{yueshu5}) indicates that the data amount of subtasks offloaded by UE $i$ is no more than the data amount of its entire task $T_i$.
\vspace{-1mm}
\section{GAME THEROY BASED PROBLEM ANALYSIS}
% In this section, we introduce the data offloading incentive to study the actual data offloading problem.
% \vspace{-1mm}
\subsection{Stackelberg Game Model}
\begin{figure}[tbp]
    \centering
    \includegraphics[scale=0.2]{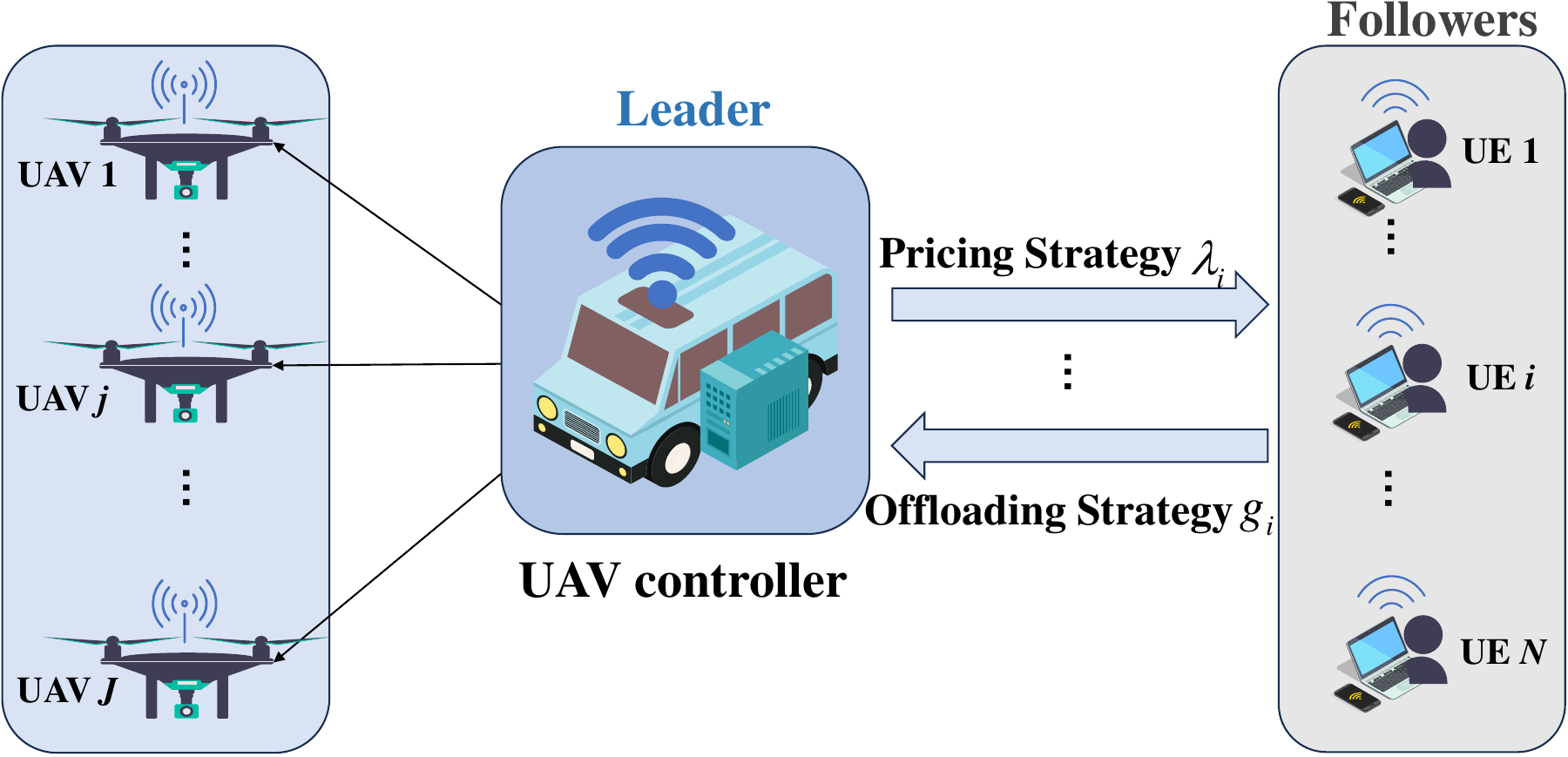}
    \caption{The Stackelberg game procedure.}
    \label{sta}
\end{figure}

% To incentivize the involvement of UAV controller in the task offloading process,
Since the UAV controller has a competitive relationship with UEs,
the interaction between UEs and the UAV controller is modeled as a Stackelberg game with a single leader and multiple followers,
as depicted in Fig. \ref{sta}.
% In this model, the UAV controller serves as the leader, setting the resource price and directing UAVs with available computational resources to assist the UEs.
% The UEs, as followers, offload a portion of their tasks to the UAVs in exchange for the corresponding compensation.
The game is described in two stages. 
In the first stage, each UE submits the task information and the UAV controller sets the resource prices $\boldsymbol{\lambda}=\{\lambda_{1},\lambda_{2},\dots,\lambda_{I}\}$ based on the information provided by the UEs.
In the second stage, each UE determines the data offloading strategy $\boldsymbol{g}=\{g_1, g_2, \dots, g_I\}$, according to the specified pricing scheme.
\vspace{-1mm}
\subsection{Optimization of UE }
We apply a backward induction method to deal with the game-theoretic problem. 
% In the first stage, we determine the optimal offloading strategy for each UE. 
% In the second stage, we derive the optimal pricing strategy of UAV controller, which depends on the offloading choices made by all UEs. 
% We also establish the existence and uniqueness of the Nash equilibrium for the problem under consideration.
The proof of the existence and uniqueness of Nash equilibrium is given as follows.
\vspace{-1mm}
\newtheorem{Definition}{\textbf{Definition}}

\begin{Definition}
    \textit{There exists Nash equilibrium among EUs with $\boldsymbol{g^{*}} = \{g_{1}^{*}, g_{2}^{*}, ..., g_{N}^{*}\}.$
    At this point, there is a utility
function $U_{i}\left(g_{i}^{*}, g_{-i}^{*}\right)>U_{i}\left(g_{i}, g_{-i}^{*}\right)$, where $g_{-i}^{*}$ is the best strategy for other UEs excluding EU i.}
\end{Definition}
\vspace{-1mm}
\newtheorem{Theorem}{\textbf{Theorem}}
\begin{Theorem}
    \textit{In the Stackelberg game, there exists a unique Nash equilibrium point when the utility function of UE $i$ adheres to Eq. (\ref{UUi}).
    In this case, the optimal offloading strategy of UE $i$ is denoted by
    \begin{equation}    g_i^*=\frac{\delta_{i}}{\frac{p_{i}}{r_{i,j}}-\varepsilon_{i}+\lambda_{i}}-1.   \end{equation}}
\end{Theorem}   
\newtheorem*{Proof}{Proof}
\vspace{-1mm}
\begin{Proof} 
    The first and second partial derivatives of the utility function $U_i$ with respect to $g_{i}$ can be obtained as follows

    \begin{equation}
        \frac{\partial U_i}{\partial g_i}=\frac{\delta_{i}}{1+g_{i}}-\frac{p_{i}}{r_{i,j}}+\varepsilon_{i}-\lambda_{i},
    \end{equation}
    and 
    \begin{equation}
        \frac{\partial U_{i}^{2}}{\partial g_{i}^{2}}=-\frac{\delta_i}{\left(1+g_i\right)^2}.
    \end{equation}

Since $\delta_i>0$ and $1+g_i>0$, we can obtain that the second derivative is less than zero. 
The utility function $U_i(g_i)$ of UE $i$ is strictly concave.
Therefore, the Nash equilibrium solution exists.
When ${\partial U_i}/{\partial g_i}=0$, we obtain the optimal data offloading strategy for UE.\qedhere
\qed
\end{Proof}
Since $g_{i}\in[0,G_{i}]$, we can obtain the threshold for the pricing strategy of the UAV controller as:
% \begin{equation}
    $\lambda_i^{min}=\frac{\delta_i}{(1+G_i)}-\frac{p_i}{r_{i,j}}+{\varepsilon_i}$,
% \end{equation}
and
% \begin{equation}
    $\lambda_i^{max}={\delta_i+\varepsilon_i}-\frac{p_i}{r_{i,j}}$.
% \end{equation}
Therefore, the optimal offloading strategy of UE $i$ can be expressed as:
\begin{equation}
    g_i^*=\begin{cases} G_i,& \lambda_i\leq\lambda_i^{min},\\
        \frac{\delta_{i}}{\frac{p_{i}}{r_{i,j}}-\varepsilon_{i}+\lambda_{i}}-1,& \lambda_i^{min}<\lambda_i<\lambda_i^{max},\\ 
         0,& \lambda_i\geq\lambda_i^{max}.\end{cases}\label{gi}
\end{equation}
\subsection{Optimization of the UAV Controller Server}

\begin{Definition}
    If \textit{$U_{con}(\lambda_{i}^{*}, g_{i}^{*}) > U^{edge}(\lambda_{i}, g_{i}^{*})$, a unique Stackelberg equilibrium is proposed between UEs and the UAV controller.
}
\end{Definition}
\vspace{-1mm}
\begin{Theorem}
    \textit{In the Stackelberg game, there exists a unique Nash equilibrium point when the utility function of UAV controller adheres to Eq. (\ref{UUcon}).
    Similarly, the pricing strategy optimal for the UAV is calculated by
    \begin{equation}    
        \lambda_i^*=\frac{\sqrt{r_{i,j}\delta_{i}(p_{i}-r_{i,j}\varepsilon_{i}+r_{i,j}{\frac{\alpha \cdot M_{j}P_j^{comp}}{f_{j}}}X_{i,j})}-p_{i}+r_{i,j}\varepsilon_{i}}{r_{i,j}}.   \end{equation}
    }
    % where $\Omega=\sqrt{r_{i,j}\delta_{i}(p_{i}-r_{i,j}\varepsilon_{i}+r_{i,j}{\frac{\alpha \cdot M_{\mathrm{uav},j}P_j^{comp}}{f_{\mathrm{uav},j}}}x_{i,j})}$.
\end{Theorem}
\vspace{-1mm}
\begin{Proof}
    By putting $g_i^*$ into $U_{con}$, we can obtain $U_{con}(\lambda_i,g_i^*)$ as
    \begin{equation}
        \begin{split}
        U_{con}(\lambda_{i},g_{i}^{*})=\sum_{i=1}^{N}\lambda_{i}(\frac{\delta_{i}r_{i,j}}{p_{i}-\varepsilon_{i}r_{i,j}+\lambda_{i}r_{i,j}}-1)\\
        -\sum_{j=1}^{J}\sum_{i=1}^{N}\frac{\alpha \cdot M_{j}}{f_{j}}(\frac{\delta_{i}r_{i,j}}{p_{i}-\varepsilon_{i}r_{i,j}+\lambda_{c}r_{i,j}}-1)X_{i,j}.
        \end{split}
    \end{equation}
    
    The first and second partial derivatives of the utility function $U_{con}$ with respect to $\lambda_{i}$ can be obtained as
   
    \begin{equation}
        \frac{\partial U_{con}}{\partial \lambda_i}=\frac{r_{i,j}\delta_{i}(p_{i}-r_{i,j}\varepsilon_{i}+r_{i,j}{\frac{\alpha \cdot M_{j}P_j^{comp}}{f_{j}}}X_{i,j})}{(p_{i}-r_{i,j}\varepsilon_{i}+r_{i,j}\lambda_{i})^{2}}-1,
    \end{equation}
and
    \begin{equation}
        \frac{\partial U_{con}^{2}}{\partial \lambda_{i}^{2}}=-\frac{2r_{i,j}^2\delta_i(p_i-r_{i,j}\varepsilon_i+r_{i,j}{\frac{\alpha \cdot M_{j}P_j^{comp}}{f_{j}}}X_{i,j})}{(p_i-r_{i,j}\varepsilon_i+r_{i,j}\lambda_i)^3}.
    \end{equation}

Since $g_i^*$ is no less than 0, we can obtain $\frac{p_{i}}{r_{i,j}}-\varepsilon_{i}+\lambda_{i}>0$. 
Therefore the second derivative is less than zero. 
When ${\partial U_{con}}/{\partial\lambda_i}=0$, we obtain the optimal pricing strategy.
\qed
\end{Proof}
\section{ALGORITHM DESIGN\label{sec:algorithm}}
In this section, we design a low computing cost mechanism ULAR to find the best location of UAV-MEC, and adjust the unloading destination of UEs according to the maximum data capacity of UAV-MEC.
Then, in order to achieve the Nash equilibrium efficiently, we propose the CPPO algorithm.
\vspace{-2mm}
% \subsection{UAVs temporary location confirmed}
\begin{algorithm}[tbp]
    \label{s1}
	\caption{ULAR Mechanism}%算法标题
	\begin{algorithmic}[1]%一行一个标行号
        \REQUIRE $\{z_1, z_2\cdots, z_I\}$, $\boldsymbol{g}$, and $J$.
        \ENSURE $\{C_1\cdots C_J\}$, $\{z_1,z_2,\cdots z_J\}$, and $X_{i,j}$.
        \STATE \textbf{Initialize} set cluster number $\{C_1\cdots C_J\}$, the maximum iterations $N$, and maximum UAV load $D_j$.\label{1}
        \STATE Select $J$ random samples from $\{z_1, z_2\cdots, z_I\}$ as the initial centroid $(\mu_1\cdots \mu_J)$.

        \FOR{$i=1$ to $N$}
        \FOR{$j=1$ to $J$}
        
        \STATE Calculate the initial distance $d_{i}^*$.
        \STATE Allocate $z_i$ to $C_j$ with minimum $d_i^j$.
        \STATE Update $\mu$ according to $\mu_1$.
        \ENDFOR

		\IF{$\mu_{j(n)}=\mu_{j(n-1)}$}
        \STATE End loop.
		\ENDIF
		\ENDFOR \label{12}
        % \STATE Receives the size of all UEs unload data $\boldsymbol{g}=\{g_{1},....,g_{i},...,g_{N}\}$;
    \WHILE{$\sum_{i\in C_j}g_i>D_j $} \label{13}
    \STATE Removal $z_i$ with $max(d_i^j)$ from $C_j$.
    \STATE Add $z_i$ to $C_{\iota}\neq C_{j}$ with minimum $d_i^j$.
    \ENDWHILE \label{16}

	\end{algorithmic}
\end{algorithm}
\vspace{-1mm}
\subsection{UAV Localization and Availability Response Mechanism}
\label{sec:kf}
% In real scenarios, the total amount of data processed by UAV-MEC at the same time is limited. 
In order to deploy UAVs and allocate computing resources of UAVs more efficiently, we design the ULAR mechanism, as shown in algorithm \ref{1}.
% The mechanism consists of the following three steps:  
% \subsubsection{step 1}
% UEs are randomly distributed in the selected square area, and UE has no other relationship with UE.
% UAVs preliminarily receive all UEs locations through preliminary scanning.
% We will solve the UAVs temporary hover position problem by K-Means algorithm.
% % This unsupervised learning approach trains without prior knowledgeand converges quickly. 
% % These are why we chose this algorithm.
% \subsubsection{step 2}
% Secondly, when the temporary hovering position of UAVs is determined, each UE makes a preliminary offloading strategy selection according to the hovering position of UAVs, and sends the value of the decided offloading data size to the selected UAV instead of the content required to be calculated.
% \subsubsection{step 3}
% When each UAV receives the value of the offloading task of UEs, it will judge whether its required total computing resources exceed its own threshold ${f}_{j},j\in\mathcal{J}$.
% That is, whether the UAV can serve all UEs simultaneously.
% If $\sum_{i\in\mathcal{N}}g_{i}X_{i,j}\leq D_j,\forall j\in\{\mathcal{J}\}$, these UAVs can provide offloading services for UEs that have selected themselves as offloading destinations.
% If $\sum_{i\in\mathcal{N}}g_{i}X_{i,j}> D_j,\forall j\in\{\mathcal{J}\}$, 
% in order to serve more UEs in this temporary emergency scenario, 
% The farthest UE in the cluster is eliminated and assigned to the next nearest cluster. 
% Repeat the process until each cluster has the right amount of UEs.
We adopt the K-means clustering based to select the optimal deployment positions of UAVs.
% Firstly, we initialize the locations of UEs, the number of clusters $J$ and the offloading strategy $\boldsymbol{g}$.
From steps \ref{1} to \ref{12}, the UEs are clustered according to the distance and the updated clusters are obtained.
Further, when the amount of offloading data exceeds the maximum load of UAV $j$, the farthest UE in the cluster assigned to the next nearest cluster from steps \ref{13} to \ref{16}. 
% Repeat the process until each cluster has the right amount of UEs.
In general, with the Algorithm 1, more UEs can be served in the limited resources of UAVs.
% Specifically, the cost only involves the cost of sending data size and allocating computing power. 
% With a small computing cost, the computing resources of UAVs can be fully utilized.

% To do this, the cluster
% partition $\{C_1,C_2,\cdots ,C_J\}$ with the lowest $\mathbb{SSE}$ in the following equation is first found using a conventional K-means approach. 
% Before assigning each UE to the closest cluster and recalculating the centers of mass for each cluster, the algorithm
% chooses $\mathbb{J}$ random UEs to serve as the initial centers of mass
% for each cluster. The algorithm repeats this operation until the
% centers of mass of every cluster remain constant.
% % \begin{equation}
%     $\mathbb{SSE}=\sum_{\iota=1}^I\sum_{q\in C_\iota}||q-\mu_\iota||^2$,
% % \end{equation}
% where the vector mean of cluster $C_l$, or the center of mass, is represented by $\mu_l$, as the following equation illustrates

% \begin{equation}
%     \mu_l=\frac{1}{|C_l|}\sum_{q\in C_l}q.
% \end{equation}

% Next, we consider the limited computational power of a single computer.
% Therefore, it is essential to implement appropriate modifications when the number of UEs in any cluster surpasses the computational load limit of the UAV. In cases where a cluster contains redundant components, the terminal located farthest from the center of its respective cluster will be removed and reassigned to the nearest available cluster. This procedure is repeated until each cluster maintains an optimal number of UEs.
\begin{algorithm}[t]

	\caption{Chess-like PSOPSSL Optimization Algorithm }%算法标题
	\begin{algorithmic}[1]%一行一个标行号
        \label{s2}
        \REQUIRE $X_{i,j}$, $\{C_1, C_2,\dots, C_J\}$, $\{z_1,z_2, \dots, z_J\}$,
        $I$, $J$, and parameters for PSOPSSL.
        \ENSURE The optimal pricing strategy $\boldsymbol{\lambda^*}$, the optimal offloading strategy $\boldsymbol{g^{*}}$, $U_{con}^{*}$, and $U_i^*$.
    % \renewcommand{\algorithmicrequire}{\textbf{Input:}}
    % \renewcommand{\algorithmicensure}{\textbf{Output:}}
        % \STATE Set $\lambda_i\left(\mathbf{0}\right)=\{\lambda_{i,1}\left(0\right),\lambda_{i,2}\left(0\right),\lambda_{i,3}\left(0\right),\lambda_{i,4}\left(0\right)\mid\forall i\in N\}$ as initial pricing decisionprofile for the UAV Controller;        
        % \STATE Set $g_i\left(\mathbf{0}\right)=\{g_{i,1}\left(0\right),g_{i,2}\left(0\right),g_{i,3}\left(0\right),g_{i,4}\left(0\right)\mid\forall i\in N\}$ as initial offloading decisionprofile for the UE $i$;        
        \REPEAT
        % \FOR{$J=1$ to $J$} 
            \FOR{$i=1$ to $I$}\label{22}
                \STATE Update pricing strategy $\lambda_i(j)$ of the the UAV controller by the PSOPSSL algorithm.
                \STATE Update offloading strategy $g_i(j)$ of the UE $i$ by using (\ref{gi}).
                %  Eq.(\ref{gi});
            \ENDFOR\label{25}
            % \IF{$\frac{|\lambda_{i}(j)-\lambda_{c}(j-1)|}{|\lambda_{i}(j)|}\leq\phi $}
            %     \STATE Break;
            % \ENDIF
        
            \STATE Generate the optimal pricing set $\lambda^{*}$=$\{\lambda_{1}^{*}, \lambda_{2}^{*}, \dots, \lambda_{I}^{*}\}$
                and the optimal offloading set $\boldsymbol{g}^{*}$=$\{g_{1}^{*}, g_{2}^{*}, \dots, g_{I}^{*}\}$.
                
        % \FOR{$j=1$ to $J$}
        %     \IF{$|C_{j}|>f_j $}
        %     \STATE Calculate the current task offloading amount of the $j$th cluster population.
        %         \STATE Removal $q_i$ with $max(d_i^j)$ from $C_k$;
        %         \STATE Add $q_i$ to $C_{\iota}\neq C_{k}$ with minimum $d_i^j$;
        %     \ENDIF
        % \ENDFOR
                
        \WHILE{$\sum_{i\in C_j}g_i^*>D_j $} \label{27}
            \STATE Remove $z_i$ with $max(d_i^j)$ from $C_j$.
            \STATE Add $z_i$ to $C_{\iota}\neq C_{j}$ with minimum $d_i^j$.
        \ENDWHILE \label{210}
    \STATE Calculate the current utility of the UAV Controller.\label{211}
    \STATE Calculate the current utility of the UE.\label{212}
        % \ENDFOR
    \UNTIL a Nash equilibrium is obtained.
        
	\end{algorithmic}
\end{algorithm}
\vspace{-2mm}
\subsection{Chess-like PSOPSSL Optimization Algorithm}
We propose the CPPO to obtain the Nash equilibrium in Algorithm \ref{s2}.
% We first initialize various parameters.
% such as the number of Uavs and users, the initial pricing strategy and offloading strategy, and the initialization parameters of PSOPSSL algorithm, etc.
% Second, The initial price strategy randomly takes 4 feasible values from the interval $[\lambda_i^{min},\lambda_i^{max}$] (line 1-2).
The particle swarm optimization probability based strategy selection learning optimization (PSOPSSL) method is used to optimize the pricing strategy and offloading strategy to find the optimal solution (lines \ref{22}-\ref{25}).
We use the ULAR mechanism to simulate a scenario similar to the two-player game in chess, with the obtained solution to avoid the situation that the UAV is not available after optimization (lines \ref{27}-\ref{210}).
The utility of both sides is calculated to reach Nash equilibrium (lines \ref{211}-\ref{212}).
The iteration is carried out until a Nash equilibrium is reached.

Inspired by learning automata theory\cite{suanfa2},  we propose the PSOPSSL method and the position update formula is
% The parameters of the standard PSO are fixed, 

\begin{equation}
    \begin{split}
    v_{ij}\left(t+1\right)=&wv_{ij}\left(t\right)+c_1^\tau r_1\left(t\right)\left[p_{ij}\left(t\right)-x_{ij}\left(t\right)\right]\\
    &c_2^\tau r_2\left(t\right)\left[p_{ij}\left(t\right)-x_{ij}\left(t\right)\right],
    \end{split}
\end{equation}
where $c_1^\tau$ is updated as $c_1^\tau=c_1^0+0.1*s_1^\tau$, and $s_1^\tau$ is the round whose local extremums are invariant.
$c_2^\tau=c_2^0+0.1*s_2^\tau$ and $s_2^\tau$ are the global extremum invariant round.
$w$ represent the weight that keeps initial velocity, which can be expressed as $w=w-0.1*s_{1}^{\tau}$.
$\tau$ is the number of iterations of PSOPSSL algorithm.
The PSOPSSL adjusts the size of its parameters based on the number of iterations and the frequency of selecting the same extreme value position. 
% This self-adjustment ensures that each particle explores the entire search space independently in the early stage, while enabling faster convergence towards optimal positions in later stages. 
It effectively balances both global and local search capabilities of particles.

\section{Simulation Result\label{sec:Simulation Result}}
In this section, we carry out the simulations using MATLAB.
The energy consumption data per unit task of UEs is set to $[0.2, 0.5] \mathrm{J/MB}$.
$\delta_i$ is set as $40$.
% The computation resource conversion $l$ is set as $2\mathrm{GHz/Mb}$.
For the UAV-MEC, the computing power is set as between [1, 5]$\times10^{9}$cycles/s.
The CPU power is a random value between [0.1, 0.5]W.
We consider a data unit requires a CPU revolution of 1,900 cycles/byte, i.e., $\alpha=1,900$ cycles/byte. 
The amount of data for each task is distributed in the range [10, 50]MB.

% In the Stackelberg game framework, the UAV controller, acting as the leader, formulates its pricing strategy by optimizing its own objectives in response to the strategies adopted by the follower agents. Subsequently, the follower UEs determine their optimal uninstall policies based on the leader's decisions ,shown in Fig.\ref{stack}.
\begin{figure}[tbp]%
    \centering
     \subfloat[Optimal offloading of the UE.]{
    \label{Ui}
    \includegraphics[width=0.40\linewidth]{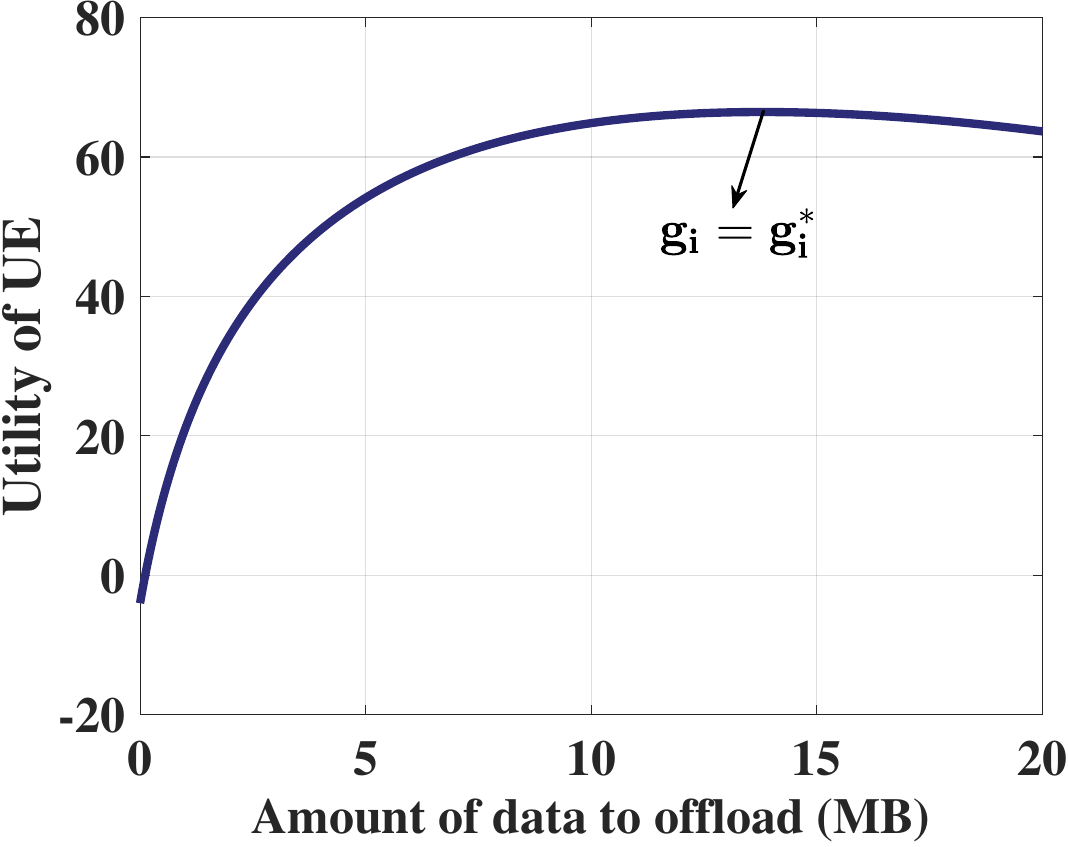}
    }\hspace{0\linewidth}
    \subfloat[Optimal price of the UAV.]{
        \label{Ucon}
        \includegraphics[width=0.40\linewidth]{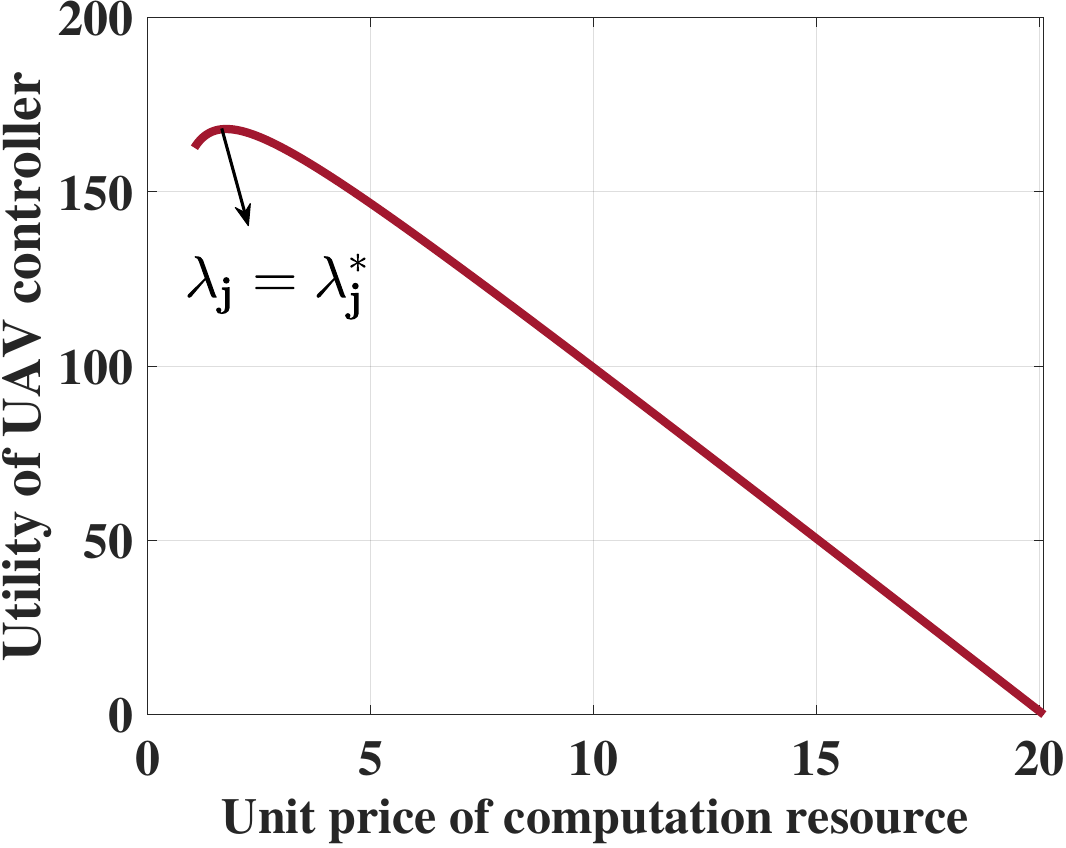}
        }
\caption{The Stackelberg equilibrium.}
\label{stack}
\end{figure}
Fig. \ref{stack} depicts the process of Stackelberg equilibrium.
Fig. \ref{stack}(a) is the utility of UE $i$ with a fixed UAV controller price $\lambda_i=\lambda_i^*$.
% The unloading policy of UE $i$ is represented on the horizontal axis, and 
The utility of UE $i$ reaches the maximum when the unloading policy is $g_i=g_i^*$. 
Fig. \ref{stack}(b) illustrates the income generated when the UAV controller adopts the fixed unloading strategy $g^*=g_i^*$. 
The optimal price $\lambda_i=\lambda_i^*$ corresponds to the point of the maximum profit. 
This analysis demonstrates that both UEs and UAV controller collaborate to achieve a Nash equilibrium.

\begin{figure}[tp]%
    \centering
    \subfloat[Utility of UE.]{
        \label{Uiy}
        \includegraphics[width=0.4\linewidth]{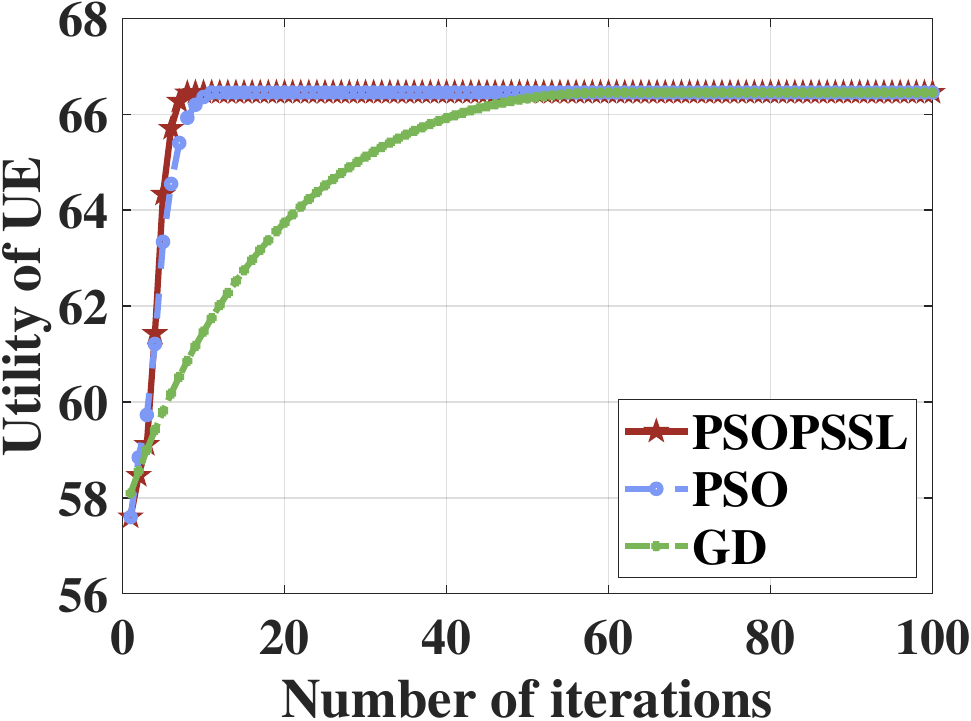}
        }\hspace{0\linewidth}
    \subfloat[Utility of UAV controller.]{
        \label{Ucony}
        \includegraphics[width=0.4\linewidth]{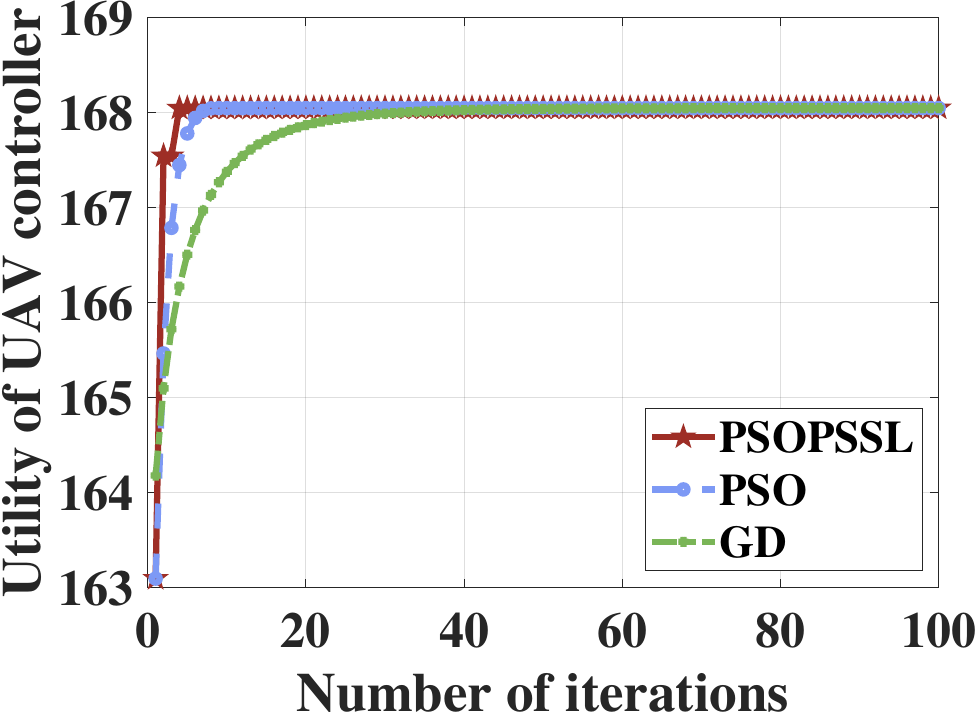}
        }
    \caption{Iterations of PSOPSSL algorithm.}
    \label{y}
\end{figure}
Fig. \ref{y} illustrates the iterative process of the proposed PSOPSSL.
Fig. \ref{y}(a) shows the iterative convergence process for the UE. 
As iterations increases, the utility progressively improves, reaching the stable convergence after 10 iterations. 
% To evaluate the performance of the PSOPSSL algorithm, the initial value is randomly selected.
Similarly, as shown in Fig. \ref{y}(b), the convergence is achieved after approximately 10 iterations as the number of iterations increases.
It is noted that PSOPSSL converges faster than particle swarm optimization (PSO) and gradient descent (GD).

\begin{figure}[tp]%
    \centering
    \subfloat[Mean utility of UEs.]{
        \label{jieguo1}
        \includegraphics[width=0.41\linewidth]{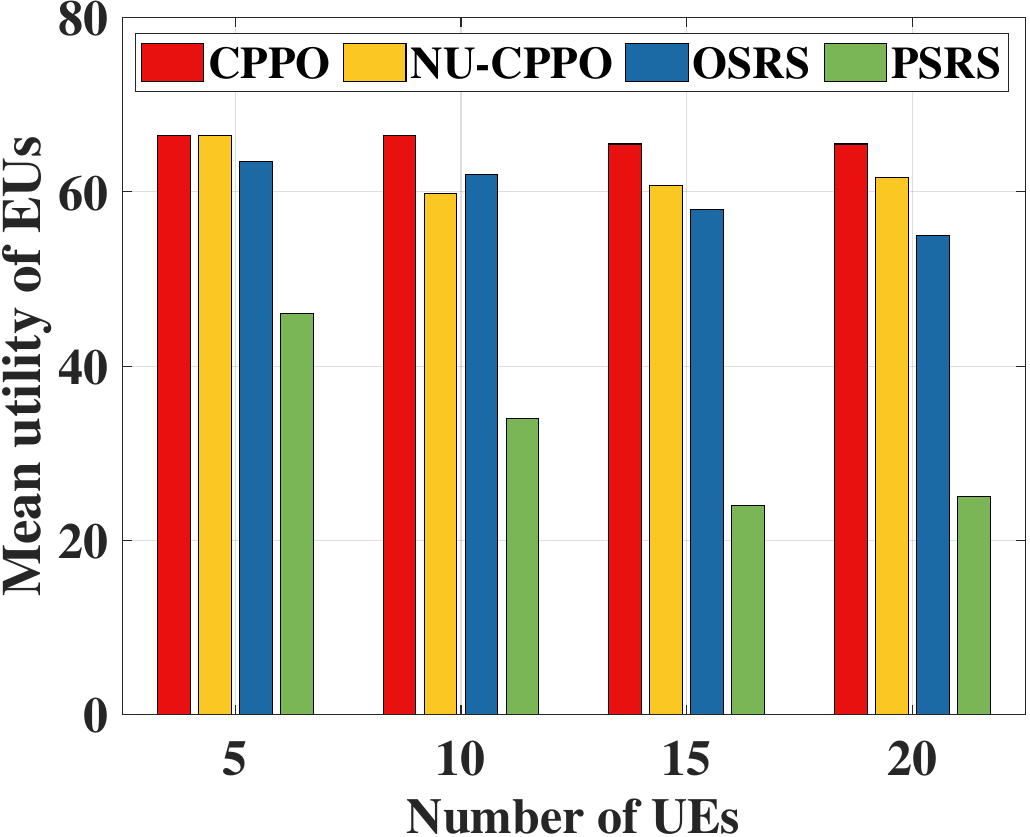}
        }\hspace{0\linewidth}
    \subfloat[Utility of UAV controller.]{
        \label{jieguo2}
        \includegraphics[width=0.41\linewidth]{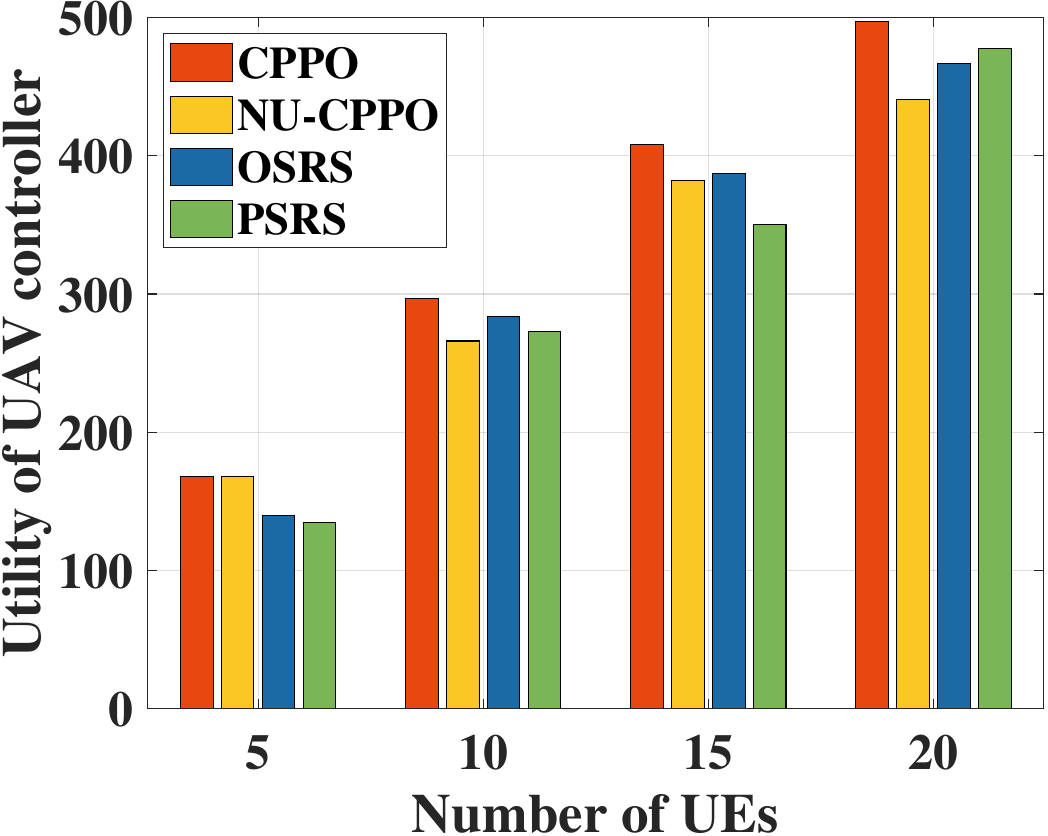}
        }
        % 0.46
    \caption{Comparison of different algorithms.}
    \label{jieguo}
\end{figure}

% Finally, to evaluate the performance of the CPPO scenario presented in this article, we will compare it with the three scenarios below.
% \subsubsection{}
% \textit{Non-ULAR mechanism CPPO strategy.(NU-CPPO):}
% First, the CCPO algorithm does not use the ULAR mechanism(NU-CPPO).
% \textit{Offloading Strategy Based on Random Selection(OSRS):}
% Second, only unloading strategies are determined using game theory methods(OSRS).
% \textit{Pricing Strategy Based on Random Selection(PSRS):}
% Third, only pricing strategies are determined using game theory methods(PSRS).
To assess the performance of the proposed CPPO method, we compare it with three alternative scenarios: 1) the CCPO algorithm without the ULAR mechanism (NU-CPPO), 2) uploading strategies determined via game theory (OSRS), and 3) pricing strategies determined via game theory (PSRS).
In detail,
Fig. \ref{jieguo}(a) and \ref{jieguo}(b) compare the mean utility of UEs for all
users and the utilize of the UAV controller under four different
offloading strategies, respectively.
% The advantages of the CPPO strategy can be seen in Figs. \ref{jieguo1} and \ref{jieguo2}.
% The CPPO algorithm outperforms the other baseline algorithms.
% Compared to other strategies, the CPPO strategy significantly improves the profitability of the UAV controller.
% In terms of the utility for UEs, the CPPO 
The CPPO scheme significantly improves the utility of UEs and the UAV controller compared to the OSRS and PSRS.
Compared to the NU-CPPO strategy, the results of the comparison highlight the necessity of implementing the ULAR mechanism. 
% As the number of UEs increases, the complexity of achieving reasonable allocation under the NU-CPPO mechanism rises, leading to a decline in its effectiveness.

% \begin{figure}[htbp]
%     \centering
%     \includegraphics[scale=0.4]{../MEC_UAV/jieguo1.eps}
%     \caption{Comparison of the mean utility of UEs under four different offloading schemes.}
%     \label{jieguo1}
% \end{figure}

% \begin{figure}[htbp]
%     \centering
%     \includegraphics[scale=0.4]{../MEC_UAV/jieguo2.eps}
%     \caption{Comparison of the mean utility of UAV Controller under four different offloading schemes.}
%     \label{jieguo2}
% \end{figure}

\section{Conclusions\label{sec:Conclusions}}
This paper presents a novel framework in which a ground-based UAV controller coordinates the deployment of UAV-MECs within 3D corridors and efficiently allocates computational resources to support damaged UEs in disaster areas.
Considering that there is a competitive relationship between the UAV controller and UEs, it is modeled as a Stackelberg game problem.
We design a ULAR mechanism to optimize UAV placement and resource allocation.  
To deal with the formulated Stackelberg game problem, we propose a CPPO algorithm.  
Extensive simulations are conducted to assess the performance of the proposed methods, with results showing that the proposed algorithms outperform other baseline approaches.
% This paper designs a novel framework that leverages multiple UAV-MECs deployed within 3D corridors to provide computation services for UEs. 
% Our approach innovatively uses a ground UAV controller to coordinate UAV-MEC deployment and allocate computational resources.
% we design a K-means based ULAR mechanism to reasonably deploy location and allocate the computing resources of UAV.
% We propose a CPPO to solve the established Stackelberg game problem.
% Simulations are conducted to evaluate the performance of the proposed methods, and the
% results demonstrate that the proposed algorithms perform better than other baseline methods.
\textcolor{black}{\bibliographystyle{IEEEtran}
\bibliography{ref}}
\end{document}